# Multi-Objective Optimization, different approach to query a database

Matteo, MC, Cordioli

Politecnico di Milano
Milan, Italy
matteo.cordioli@mail.polimi.it

**Abstract**

The datasets available nowadays are very rich and complex, but how do we reach the information we are looking for? In this survey, two different approaches to query a dataset are analyzed and algorithms for each type are explained. Specifically, the TA and NRA have been analyzed for the Top-K query and the Basic Block-Nested-Loops has been examined for the skyline query. Moreover, it's explained the core idea behind the Prioritized and Flexible skyline. In the end, the pros and cons of each type of analyzed query have been evaluated based on different criteria.

## 1 Introduction

Answering with exact data to an interrogation in a modern dataset is a complex procedure, and most of the time it's impossible. Users often query for specific details, and they expect a reasonable answer in a short amount of time. Many times, even if we are used to computers with high performances, the query is too heavy to be computed in an exact form or sometimes even in any form at all. Modern datasets try to answer our interrogation with the closest data they can reach that matches our request. But what does it mean "closest data"? The tuples that according to a scoring function or a dominance constrain are considered the best.

This paper considers two different approaches:
- Ranking queries (or Top-K): the best K tuples are computed on the base of a single-objective function where we consider many different parameters weighted on a scoring function and a user-given prioritization.
- Skyline: the answers are the non-dominated tuples. We will consider the traditional skyline and the (notion of) flexible skyline

After introducing these two different approaches in more detail, this article aims to compare them.

## 2 Ranking Queries (Top-K)

In many cases, the users are interested in the (K) most fitting query in the dataset they are working with. One possible way to find out the K tuples is by processing each one and assigning a value, which we call that value score. A score is usually assigned by a scoring function which is an aggregate of partial scores based on the user's request.

Example 2.1: *Consider a website that will provide to users the best hotels based on distance from the city center (in Km) and/or airport (in Km). An external tool will calculate the distance from the city center for each hotel in our dataset. The site will offer 3 different options of query: city center as the only constrain and so ignoring the airport, vice versa airport as the only constrain and so ignoring the city center and last the user can choose, in percentage, how much each option is important (a is the percentage for DistanceFromAirport and c for DistanceFromCenter).*

Table 1

| IdHotel | Name | DistanceFromAirport | DistanceFromCenter |
|---|---|---|---|
| 1 | Ginse | 10 | 4 |
| 2 | Firse | 18 | 5 |
| 3 | Parso | 22 | 1 |
| 4 | Nerso | 2 | 8 |

Example 2.1 is a typical problem requiring a scoring function to be solved in a modular way. We can call $F$ our scoring function and "assemble" out functions based on our needs. The first scenario explained in Example 2.1 required the easier scoring function we can write $F_1=DistanceFromCenter$ and the same is true for the second scenario where our $F$ will be computed as $F_2= DistanceFromAirport$, the third scenario required two more inputs from user $p$ and $d$ and with those we can write $F_3=a*DistanceFromAirport+c*DistanceFromCenter$ (we consider $a=0.3$ and $c=0.7$) computing these 3 different scoring functions and sorting the tuples we obtain:

Table 2

| IdHotel | Name | $F_1$ |
|---|---|---|
| 4 | Nerso | 8 |
| 2 | Firse | 5 |
| 1 | Ginse | 4 |
| 3 | Parso | 1 |

Table 3

| IdHotel | Name | $F_2$ |
|---|---|---|
| 3 | Parso | 22 |
| 2 | Firse | 18 |
| 1 | Ginse | 10 |
| 4 | Nerso | 2 |

Table 4

| IdHotel | Name | $F_3$ |
|---|---|---|
| 2 | Firse | 8.9 |
| 3 | Parso | 7.3 |
| 4 | Nerso | 6.2 |
| 1 | Ginse | 5.8 |

As we can see in tables 2-3-4 (which are based on different scoring functions), we obtain three different top-4 results from the same dataset. Note that we can affirm that only post-processing all the tuples with all the different scoring functions we are analyzing.

We have scanned sequentially all database objects computing the score of each object according to each feature and combined them into a final score for each tuple. This approach suffers from scalability problems concerning database size and the number of options/features.

### 2.1 Data Access and Implementation Level Dimension

Many Top-K queries need to aggregate multiple parameters, but we face the problem of accessing all these fields to compute the final score. Lists of objects could be accessed sequentially in score order, we refer to this method as *Sorted Access*. Otherwise, we could know in advance which object is needed to compute the final score, in this case, we can directly access the right object, we refer to this method as *Random Access*. The processing techniques are classified according to which method is available to access data (not all the methods are reported):

- Sorted and random access: both accesses are available in the current DBMS to process top-k queries. The Threshold Algorithm is an example that uses both methods [2.3.1].

- No random access: only the sorted access method is available to process top-k queries. The No Random Access Algorithm is an example that uses this method [2.3.2].

**2.2 Ranking functions**

Many top-k queries algorithms use the computation of upper bound objects scores to process data and find the K ones. This action is simplified if the scoring function is monotone. A function F, defined on multiple predicates, is monotone if $F(p_1,..., p_n) \leq F(p_1',...., p_n')$ whenever $p_i < p_i'$ for every $i$.

Most of the current algorithms assume to compute the query using a monotone ranking function because the monotone property generally improves the computation time.

Some new techniques use a generic form for the scoring function, this method increases the complexity of the scoring function. Suppose to have a non-monotone function, using the bound method becomes impossible because you can't have the upper or lower bound of your function. Some recent ideas support arbitrary ranking functions by modeling top-k queries as an optimization problem [Zhang et al. 2006].

**2.3 Algorithm**

Let's now consider some classical top-k query algorithms without entering into the dataflow but trying to understand how they work. For each algorithm, we will examine the pros and cons.

2.3.1 Threshold Algorithm (TA)

Suppose to have a table where we are looking for top-k queries based on multiple predicates. First, we order the tuples by a predicate. TA will scan with sorted access based on the chosen predicate and will look for the same object with random access in the list of the other predicates. Once all the predicate scores are computed, it's possible to compute the total score of current tuples. An upper bound $T$ is maintained for the overall score of unseen objects. All the objects with total scores that are greater than or equal to $T$ can be reported as top-k solutions.

TA assumes that the cost of the two different access methods is the same, and moreover, it doesn't have a restriction on the number of random accesses to be performed during its execution.

Example 2.2: *consider the table of example 2.1 and assume as predicate one DistanceFromAirpot and as predicate two DistanceFromCenter. The first 2 steps of TA (with scoring function $F=p_1+p_2$) are shown in steps 1 and 2.*

Step 1:   T=30

| | IdHotel | p1 | | IdHotel | p2 | | Buffer |
|---|---|---|---|---|---|---|---|
| S.A-> | 3 | 22 | S.A-> | 4 | 8 | | 3 (23) |
| | 2 | 18 | | 2 | 5 | | 4 (10) |
| | 1 | 10 | | 1 | 4 | | |
| R.A-> | 4 | 2 | R.A-> | 3 | 1 | | |

Step 2:   T=23

| | IdHotel | p1 | | IdHotel | p2 | | Buffer |
|---|---|---|---|---|---|---|---|
| | 3 | 22 | | 4 | 8 | | 3 (23) |
| S.A/R.A-> | 2 | 18 | S.A/R.A-> | 2 | 5 | | 2 (23) |
| | 1 | 10 | | 1 | 4 | | 4 (10) |
| | 4 | 2 | | 3 | 1 | | |

### 2.3.2 No Random-Access Algorithm (NRA)

NRA will compute the top-k query with only sorted access and will use bounds computed over the objects exact scores. Suppose having a table with multiple predicates and the total scores are calculated as their sum. NRA starts from the top, analyzes the first object for each predicate, and calculates the upper and lower bound of that object. The upper bound is calculated by applying the scoring function (sum in our example) on the current object and the values that are still not visited are considered maximized by the last visited object. The lower bound is calculated by applying the scoring function (sum in our example) on the current object and the values that are still not visited are considered minimized by the last visited object. Once the lower bound equals the upper bound, we consider the object fully visited and we insert it in the final list of tuples; when the list contains K tuples, we reach our scope.

Example 2.3: *consider the table of example 2.1 and assume as predicate one DistanceFromAirpot and as predicate two DistanceFromCenter. The first 2 steps of NRA (with scoring function $F=p_1+p_2$) are shown in steps 1, 2, and 3.*

Step 1:

| | IdHotel | p1 | | IdHotel | p2 | | Buffer |
|---|---|---|---|---|---|---|---|
| S.A-> | 3 | 22 | S.A-> | 4 | 8 | | 3 (22-30) |
| | 2 | 18 | | 2 | 5 | | 4 (8-30) |
| | 1 | 10 | | 1 | 4 | | |
| | 4 | 2 | | 3 | 1 | | |

Step 2:

| | IdHotel | p1 | | IdHotel | p2 | | Buffer |
|---|---|---|---|---|---|---|---|
| | 3 | 22 | | 4 | 8 | | 2 (23-23) |
| S.A-> | 2 | 18 | S.A-> | 2 | 5 | | 3 (22-27) |
| | 1 | 10 | | 1 | 4 | | 4 (8-26) |
| | 4 | 2 | | 3 | 1 | | |

Step 3:

| | IdHotel | p1 | | IdHotel | p2 | | Buffer |
|---|---|---|---|---|---|---|---|
| | 3 | 22 | | 4 | 8 | | 2 (23-23) |
| | 2 | 18 | | 2 | 5 | | 3 (22-26) |
| S.A-> | 1 | 10 | S.A-> | 1 | 4 | | 1 (14-14) |
| | 4 | 2 | | 3 | 1 | | 4 (8-18) |

## 3 Skyline queries

Skyline queries are a valid alternative to top-k queries to discover interesting data in a dataset. Skylines are simple to specify, they just return all the tuples non-dominated by any others, however, they haven't any means to accommodate user's requests (preferences) or to control the cardinality of the result set. Before formally defining a skyline we must introduce dominance: one point dominates another if it is at least as desirable in every dimension, and strictly more desirable in at least one. From a user's point of view there is no support for understanding why data points are dominated, nor by how much so the differences can be quite perplexing, especially when comparing several similar queries. [4]

In this section one implementation of the algorithm to skyline query will be analyzed alongside two different types of skyline queries prioritized skyline, flexible skyline.

## 3.1 Algorithm and Implementation

The native way to compute the Skyline is to apply a nested-loops algorithm and compare each tuple to every other tuple. Obviously, this is very inefficient.

The Basic Block-nested-loops Algorithm [6] is significantly faster: rather than considering a tuple at a time, this algorithm produces a block of Skyline tuples at every iteration.

The idea of this algorithm is to keep a buffer of incomparable tuples in the main memory. When a tuple p is read from the input, p is compared to all tuples of the buffer, and based on this comparison, p is either eliminated, or placed into the buffer, or into a temporary file which will be considered in the next iteration of the algorithm. Only three cases can occur:
- p is dominated by a tuple within the buffer, p is eliminated, and will not be considered in future iterations.
- p dominates one or more tuples in the buffer, these tuples are eliminated and will not be considered in future iterations and p is stored in the buffer.
- p is incomparable with each tuple in the buffer, if the buffer has some more space p is inserted otherwise p is written to a temporary file on disk.

At the end of each iteration, we can output tuples of the buffer which have been compared to all tuples that have been written to the temporary file; these tuples are not dominated by other tuples, and they do not dominate any tuples which are still under consideration.

This algorithm works particularly well if the skyline is "small". In the best-case complexity is of the order of $O(n)$: n being the number of tuples in the input. In the worst case, the complexity is of the order of $O(n^2)$, the asymptotic complexity is the same as for the native nested loops algorithm, but the block-nested-loops algorithm shows much better I/O behavior.

## 3.2 Prioritized Skyline

Prioritized Skyline, or p-skyline, queries are a generalization of skyline queries in which the user can specify that some attributes are more important than others, by respecting the syntax of so-called p-expressions. A p-expression will have fewer than d "most important" attributes, where d is the number of attributes the expression is working with. Therefore p-skylines usually contain fewer tuples than skylines.

To construct a p-skyline relation from a skyline relation, one needs to provide a p-graph describing relative attribute importance. A p-graph is a graph whose nodes are attributes and edges go from more to less important attributes, it satisfies the properties of transitivity and irreflexivity. However, requiring users to explicitly describe attribute importance seems impractical for several reasons, mainly the number of pairwise attribute comparisons required may be not fully aware of their own preferences. To address this problem Denis Mindolin and Jan Chomicki [5] developed a method of elicitation of p-skyline relation based on simple user-provided feedback: given a set of tuples the user must identify which he/she likes and dislikes most.

### 3.2.1 Properties

In this section are listed some of the fundamental properties of p-skyline relations [5]:
- Every p-skyline relation the winnow query result will always be contained in the corresponding skyline. In real life that means that if user preferences are modeled as a p-skyline relation instead of a skyline relation, the size of the query result will not be larger than the size of the skyline and it may be smaller.
- To check the equality of 2 or more p-skyline, one only needs to compare their p-graphs.
- More preference relations imply more dominated tuples and fewer most preferred ones.

## 3.3 Flexible Skyline

F-skylines are able to weigh the different importance of different attributes. However, unlike p-skylines, where a strict priority between attributes is assumed, f-skylines can model user preferences using constraints on the weights used in a scoring function, thus allowing for much greater flexibility. F-skylines use the notion of *F-dominance*: tuple t *F*-dominates tuple s when t is always better than or equal to s according to all the scoring functions in *F*. To analyze this category of queries we will take into consideration two operators introduced by Paolo Ciaccia and Davide Martinenghi [3]: ND, characterizing the set of *non-F-dominated* tuples and PO, referring to the tuples that are *potentially optimal.* We must keep in mind that ND and PO coincide with the traditional skyline when *F* is the family of all monotone scoring functions.

### 3.3.1 Non-F-Dominated, ND operator

Sorting the dataset is always beneficial to compute ND, as stated by research [3]. Computing ND starting from the skyline does not pay off in challenging scenarios. Moreover, if the set of constraints has a reasonably high selectivity, interleaving dominance, and F-dominance test leads to better performance than first checking dominance on all tuples and then F-dominance on the remaining tuples.

### 3.3.2 Potentially Optimal, PO operator

Knowing that for any set F of monotone scoring functions the following relationship holds:

$$PO \subset ND \subset Skyline$$

we can affirm that computing PO starting from ND is always the best alternative. The Potentially Optimal operator will compute the best option according to some function or to a strictly better one than all the others according to at least one monotone scoring function.

# 4 Pro and Cons Skyline – Top-k

The following table offers an overview of Top-K vs Skyline queries, considering some basic properties: simplicity of formulation, view of interesting results, control of result cardinality, trade-off among attributes [3].

| Query Type | Simplicity of Formulation | Interesting Results | Cardinality Control | Trade Attributes |
|---|---|---|---|---|
| **Top-K** | No | No | Yes | Yes |
| **Skyline** | Yes | Yes | No | No |
| **Prioritized** | Yes | Yes | No | Yes |
| **Flexible** | Yes | Yes | No | Yes |

# 5 Conclusion

Today's world is overflown by data and we are learning how to collect, analyze and use all those data, furthermore we are researching and developing different solutions to improve our query skills. One of the main problems we face today is that we need to extract/process reliable data in a short amount of time: bank transactions and web researches are typical examples of areas where we expect live feedback. Feasible solutions are available thanks to the latest technologies and algorithms. Top-k and Skyline queries are starting points for ranking data, further research could focus on different search parameters for Top-k and Skyline queries or could even try to find out new methods to query a dataset.